# Stability, Reactivity, and Aromaticity of Compounds of a Multivalent Superatom

**Pratim Kumar Chattaraj** [*] **and Santanab Giri**
*Department of Chemistry, Indian Institute of Technology, Kharagpur 721302, India*
E mail: pkc@chem.iitkgp.ernet.in

In this communication we analyze the stability, reactivity and possible aromatic behavior of two recently reported clusters (Reveles, J. U.; Khanna, S. N.; Roach, P. J.; Castleman, A. W. jr. *Proc. Natl. Acad. Sci*, **2006**, *103*, 18405), viz., $Al_7C^-$ and $Al_7O^-$ in the light of the principles of the maximum hardness and minimum electrophilicity as well as the nucleus independent chemical shift values.

In a recent issue of Chem. & Engg. News [1] it is highlighted that an $Al_7^-$ cluster mimics the behavior of a single multivalent germanium atom. S. N. Khanna and his group [2] have been involved in an important area of research involving the metal clusters visualized as super atoms, e.g. $Al_{13}$, $Al_{13}^-$ and $Al_{14}^{2+}$ clusters possess characteristics akin to that of halogen, noble gas and alkaline earth atoms respectively. A multivalent superatom, $Al_7^-$ forms stable compounds like $Al_7C^-$ and $Al_7O^-$ whose stability mimicks that of SiC and CO respectively through an appropriate shell filling as in the standard *aufbau prinzip*. $Al_7C^-$ also forms ionic compounds with alkali metals with hardly any distortion in the original cluster unit. They have adopted [2] a joint experimental- theoretical approach to synthesize these clusters and to study their properties including the gap between the associated frontier orbitals as is the standard practice in the metal cluster studies.

In the present communication we analyze the exceptional stability of these species using the electronic structure principles and the nucleus independent chemical shift calculated at the ring center [3], NICS(0) which is an indicator of the aromatic/antiaromatic behavior.

Conceptual density functional theory [4] provides definitions of global descriptors like electronegativity [5] ($\chi$), hardness [6]($\eta$) and electrophilicity [7]($\omega$) as:

$$\chi = -\left(\partial E / \partial N\right)_{v(\vec{r})} \quad (1)$$

$$\eta = \frac{1}{2}\left(\partial^2 E / \partial N^2\right)_{v(\vec{r})} \quad (2)$$

and $\omega = \chi^2 / 2\eta$ (3)

as well as local descriptors like the Fukui function [8]($f_k^\alpha$) and the philicity [9]($\omega_k^\alpha$) in terms of the respective electronic population $p_k$ at the atom $k$ as:

$f_k^+ = p_k(N+1) - p_k(N)$
$\qquad$ for nucleophilic attack $\qquad$ (4a)
$f_k^- = p_k(N) - p_k(N-1)$
$\qquad$ for electrophilic attack $\qquad$ (4b)
$f_k^0 = \frac{1}{2}(f_k^+ + f_k^-)$
$\qquad$ for radical attack $\qquad$ (4c)

and $\omega_k^\alpha = \omega \cdot f_k^\alpha$ , $\alpha = +, -, 0$ denotes nucleophilic, electrophilic and radical attacks respectively.

For a stable system or a favorable process the hardness often becomes the maximum [10] and the electrophilicity becomes the minimum [11] in most cases. In order to test the validity of these principles vis-à-vis the exceptional stability of $Al_7C^-$ and $Al_7O^-$ we calculate various global and local descriptors at the B3LYP/6-311+G** level of theory. The global quantities are calculated using a finite difference approximation and the Koopmans' theorem. Necessary charges are obtained using a natural population analysis (NPA) scheme. The aromatic behavior is analyzed using the NICS (0) values.

Figure 1 and Table 1 provide the optimized geometries of $Al_7C^-$ and $Al_7O^-$. All the systems studied here correspond to minimum energy structures on the potential energy surface as authenticated by the number of imaginary frequency to be zero. While C is endohedral in $Al_7C^-$ and, O is exohedral in $Al_7O^-$.

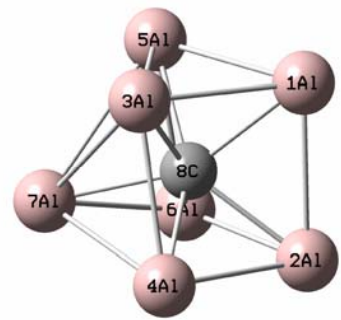

(a)

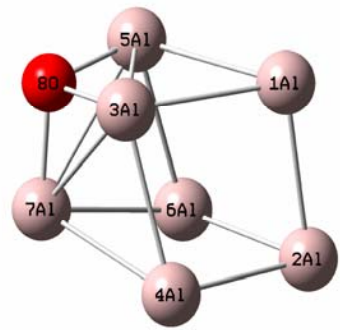

(b)

Figure 1: Optimized geometries (B3LYP/6-311+G**) of (a) $Al_7C^-$ and (b) $Al_7O^-$

Table 1: Selected geometrical parameters (bond lengths) of $Al_7C^-$ and $Al_7O^-$

| $Al_7C^-$ | | $Al_7O^-$ | |
|---|---|---|---|
| R(1,2) | 2.626 | R(1,2) | 2.646 |
| R(1,3) | 2.668 | R(1,3) | 2.745 |
| R(1,4) | 3.588 | R(1,5) | 2.735 |
| R(1,5) | 2.668 | R(2,4) | 2.634 |
| R(1,8) | 2.136 | R(2,6) | 2.639 |
| R(2,4) | 2.626 | R(3,4) | 2.736 |
| R(2,6) | 2.629 | R(3,5) | 2.813 |
| R(3,4) | 2.668 | R(3,7) | 2.821 |
| R(3,5) | 2.849 | R(3,8) | 1.886 |
| R(3,8) | 2.136 | R(4,7) | 2.753 |
| R(4,5) | 4.163 | R(5,6) | 2.736 |
| R(4,6) | 3.585 | R(5,8) | 1.888 |
| R(4,7) | 2.668 | R(6,7) | 2.749 |
| R(4,8) | 2.137 | R(7,8) | 1.884 |
| R(5,6) | 2.673 | | |
| R(5,7) | 2.849 | | |
| R(5,8) | 2.132 | | |
| R(6,7) | 2.668 | | |
| R(6,8) | 2.135 | | |

Energy (E), frontier orbital energies ($E_{HOMO}$, $E_{LUMO}$), electronegativity ($\chi$), hardness ($\eta$) and electrophilicity ($\omega$) values for X($Al_7C^-$), Y($Al_7O^-$) and their ions are provided in Table 2. Stability of X and Y is clearly delineated through their E, $\eta$ and $\omega$ values. While E and $\omega$ values of X (Y) are the lowest the $\eta$ value is the highest when compared to those values of $X^{\pm}$ ($Y^{\pm}$) as predicted by the principles of minimum energy and electrophilicity and maximum hardness. In order to check the corresponding $\Delta$ SCF values (without using Koopmans' approximation) we found that ($\eta$, $\omega$) values for X [Y] are (1.315, 0.030) [(1.106, 0.014)] which are comparable to those values reported in Table 2.

Table 2: Total energy (E,au), frontier molecular orbital energies ($E_{HOMO}$, $E_{LUMO}$, au), hardness($\eta$,ev), electronegativity($\chi$,ev) and electrophilicity($\omega$,ev) of $Al_7C^-$ and $Al_7O^-$ and their ions

| Cluster | E | $E_{HOMO}$ | $E_{LUMO}$ | $\eta$ | $\chi$ | $\omega$ |
|---|---|---|---|---|---|---|
| $Al_7C$ | -1735.246 | -0.207 | -0.139 | 0.034 | 0.173 | 0.435 |
| $Al_7C^-$ | -1735.364 | -0.067 | 0.030 | 1.318 | 0.019 | 0.100 |
| $Al_7C^{2-}$ | -1735.288 | 0.091 | 0.165 | 0.037 | -0.128 | 0.222 |
| $Al_7O$ | -1772.483 | -0.189 | -0.125 | 0.032 | 0.157 | 0.384 |
| $Al_7O^-$ | -1772.577 | -0.047 | 0.026 | 0.993 | 0.010 | 0.037 |
| $Al_7O^{2-}$ | -1772.509 | 0.092 | 0.140 | 0.024 | -0.116 | 0.282 |

The NICS (0) values associated with various rings of $Al_7C^-$ and $Al_7O^-$ are presented in Table 3. Corresponding large negative values

Table 3: Nucleus independent chemical shift (NICS (0),ppm) values of various rings in $Al_7C^-$ and $Al_7O^-$

| Cluster | Ring 1 2 4 3 | Ring 4 2 6 7 | Ring 1 2 6 5 |
|---|---|---|---|
| $Al_7C^-$ | -66.663 | -66.664 | -66.728 |
| $Al_7O^-$ | -43.058 | -42.981 | -43.067 |
| Cluster | Ring 1 3 5 | Ring 3 5 7 | Ring 5 6 7 | Ring 3 4 7 |
| $Al_7C^-$ | -54.993 | -52.973 | -54.988 | - |
| $Al_7O^-$ | -38.423 | -45.411 | -35.955 | -36.875 |

(NICS (0) value of benzene is -9.7) indicate [3] the highly aromatic nature of these clusters which is expected from such a stable ring compound.

The NPA charges and philicities at various atomic centers of $Al_7C^-$ and $Al_7O^-$ are presented in Table 4. In $Al_7C^-$ all Al atoms are preferred sites for attack by an anion or a

Table 4: Charges ($q_k$(NPA),au) and philicity($\omega_k^\alpha$,ev) on various atoms in $Al_7C^-$ and $Al_7O^-$

| Atom | No | $q_k$ | $\omega_k^+$ | $\omega_k^-$ | $\omega_k^o$ |
|---|---|---|---|---|---|
| Al | 1 | 0.291 | 0.012 | 0.017 | 0.014 |
| Al | 2 | 0.191 | 0.019 | 0.019 | 0.019 |
| Al | 3 | 0.295 | 0.018 | 0.011 | 0.015 |
| Al | 4 | 0.289 | 0.012 | 0.015 | 0.014 |
| Al | 5 | 0.297 | 0.019 | 0.013 | 0.016 |
| Al | 6 | 0.294 | 0.012 | 0.017 | 0.014 |
| Al | 7 | 0.295 | 0.019 | 0.012 | 0.015 |
| C | 8 | -2.954 | -0.010 | -0.003 | -0.007 |
| Atom | No | $q_k$ | $\omega_k^+$ | $\omega_k^-$ | $\omega_k^o$ |
| Al | 1 | -0.192 | 0.008 | 0.009 | 0.009 |
| Al | 2 | -0.269 | 0.003 | 0.001 | 0.002 |
| Al | 3 | 0.481 | 0.004 | 0.002 | 0.003 |
| Al | 4 | -0.199 | 0.004 | 0.009 | 0.007 |
| Al | 5 | 0.472 | 0.001 | 0.002 | 0.002 |
| Al | 6 | -0.195 | 0.006 | 0.009 | 0.008 |
| Al | 7 | 0.498 | 0.010 | 0.002 | 0.006 |
| O | 8 | -1.596 | -8E-05 | 0.002 | 7E-04 |

hard nucleophile [12] while the C- center is apt for an attack by a cation or a hard electrophile. The O- center and Al (1, 2, 4, 6)

atoms are good for attack by a cation / hard electrophile where as the rest of the Al atoms in $Al_7O^-$ are appropriate for the attack by an anion / a hard nucleophile. The fact is corroborated by the respective plots of the highest occupied and lowest unoccupied molecular orbitals (Figure 2). Philicity values

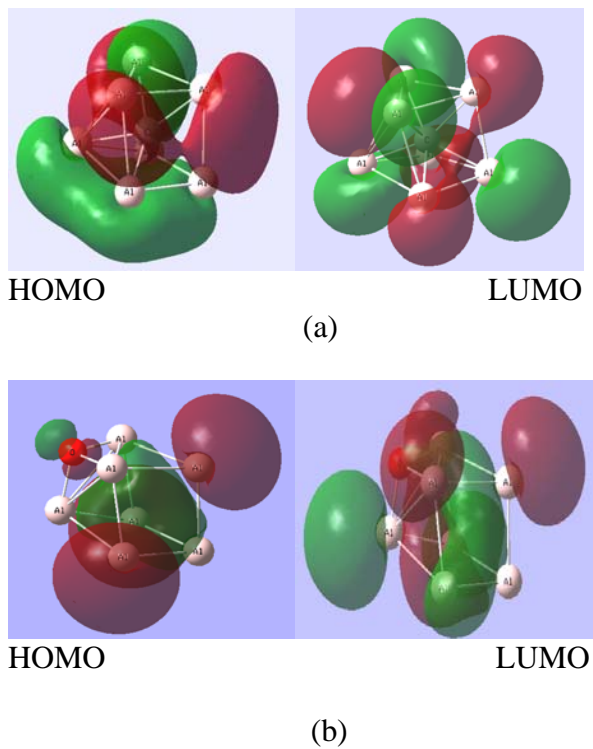

HOMO      LUMO
(a)

HOMO      LUMO
(b)

Figure 2: Frontier molecular orbital pictures of (a) $Al_7C^-$ and (b) $Al_7O^-$

provide the corresponding information regarding the attack by a soft species. $Al_7C^-$ is more electrophilic than $Al_7O^-$ since former has a larger $\omega$ value. Although C and O centers are suitable for attack by a hard electrophile those centers are bad as far as the attack of the soft electrophiles are concerned. Sites of preference for hard and soft nucleophiles are also not always same. On an average atomic sites in $Al_7C^-$ are more reactive than those of $Al_7O^-$.

In summary, both $Al_7C^-$ and $Al_7O^-$ are stable as dictated by the maximum hardness principle and the minimum electrophilicity principle. The NICS (0) values suggest their strong aromatic character. Their site selectivity towards attack by ions and hard / soft electro(nucleo) philes are analyzed in terms of atomic charges and philicities.

**Acknowledgment:** We thank BRNS, Mumbai for financial assistance.

# Table of Contents (TOC) Graphic

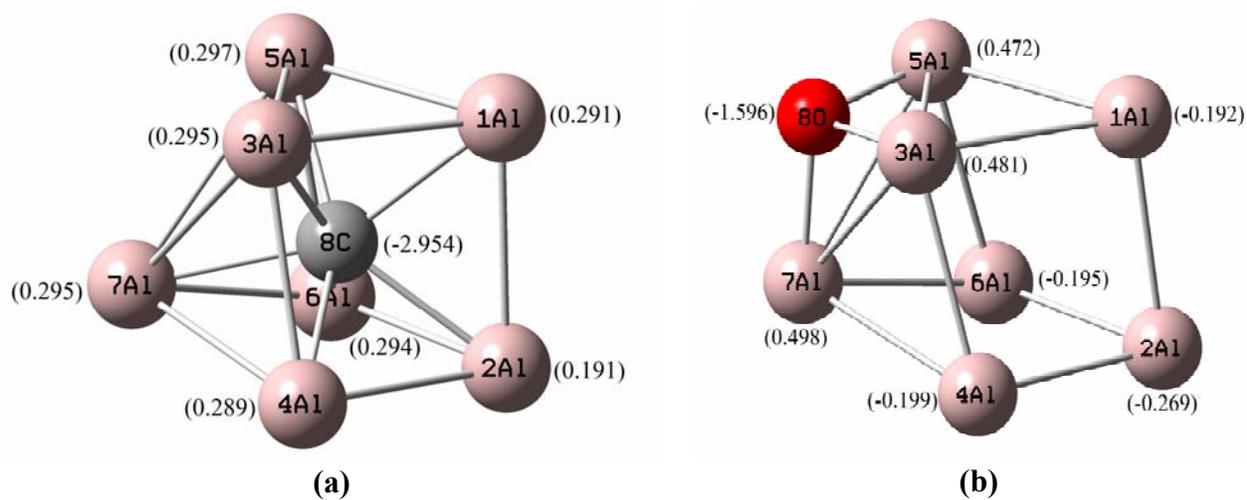

(a)             (b)

Optimized geometries (B3LYP/6-311+G**) and charges of (a) $Al_7C^-$ and (b) $Al_7O^-$

# Abstract

In this communication we analyze the stability, reactivity and possible aromatic behavior of two recently reported clusters (Reveles, J. U.; Khanna, S. N.; Roach, P. J.; Castleman, A. W. jr. *Proc. Natl. Acad. Sci*, **2006**, *103*, 18405), viz., $Al_7C^-$ and $Al_7O^-$ in the light of the principles of the maximum hardness and minimum electrophilicity as well as the nucleus independent chemical shift values.

# Keywords